\documentclass[final,1p,times]{elsarticle} 

\usepackage{graphicx}
\usepackage{amssymb} 
\usepackage{amsthm} 

\journal{Nuclear Physics A} 

\begin{document}

\begin{frontmatter} 

\title{Introductory Overview of Quark Matter 2012}

\author{Urs Achim Wiedemann}
\address{CERN PH-TH Department, CH-1211 Geneva 23}

\begin{abstract} 
 The two arguably most generic phenomena seen in ultra-relativistic heavy ion collisions are the flow of essentially all soft hadronic observables and the quenching of
 essentially all hard hadronic observables. Limiting the discussion to these two classes
 of phenomena, I review what can be said so far about the properties of hot and dense
 QCD matter from the heavy ion programs at RHIC and at the LHC, and I discuss the opportunities for further progress in the coming years. \end{abstract} 

\end{frontmatter} 


\section{Introduction}
Over the last decade, the heavy ion programs at RHIC and - more
recently - at the LHC have furnished an unprecedented amount and quality of 
information about ultra-relativistic nucleus-nucleus collisions. The main
motivation for these programmes is that they provide unique experimental access
to matter under conditions of extreme temperature and density. Data provide since 
long~\cite{Adams:2005dq,Adcox:2004mh,Heinz:2000bk}
overwhelming evidence that the energy density attained in ultra-relativistic 
heavy ion collisions is initially higher than that needed to access the partonic 
high-temperature phase of QCD, that the systems produced in these collisions
expand over time scales $O(10 {\rm fm/c})$ that are much longer than the 
characteristic time scales of individual QCD processes, and that during this long
time collective phenomena and medium effects develop that are numerically large
and thus accessible to detailed experimentation. The organizers of this Quark Matter 
Conference asked me to discuss what we have learnt so far from these data 
about the {\it fundamental} properties of hot and dense matter, and what we 
can learn with the ongoing experimental programs in the coming years. Here,
I address this question by focussing only on the two arguably most generic experimental manifestations of hot and dense matter in heavy ion collisions: the {\it flow} seen in 
essentially all soft (i.e., low-$p_T$) observables 
and the {\it quenching} of essentially all hard hadronic observables.

\section{The Panta Rhei of soft physics}
In general, the notion 'flow' refers to the correlation between the 
{\it spatial positions} from which particles are emitted and the {\it momenta} 
with which they
are emitted. But direct experimental access to spatial positions in the
overlap of two colliding nuclei is scarce. As a consequence, the standard
fluid- or aero-dynamical characterization of flow in terms of 
position-dependent averages of flow fields (such as $\langle \vec{u}(\vec{x})\rangle$,
$\langle \vec{u}(\vec{x}_1)\,  \vec{u}(\vec{x}_2)\rangle$ etc.) is of limited use.
Rather, the most direct experimental characterization of flow in heavy ion collisions
is in terms of correlations of the produced particle momenta with the global 
spatial orientation of the overlap of two colliding nuclei, i.e., with its reaction
plane(s). These correlations are characterized by the Fourier coefficients $v_n$ of
the harmonic decomposition of particle spectra in the azimuthal angle with respect to 
the reaction plane. In heavy ion collisions, the study of the $v_n$'s and fluid dynamic 
concepts have a long track-record
of successful qualitative predictions. An early success is e.g. Olitrault's argument that 
with increasing center of mass energy, the second harmonics of the azimuthal 
particle distribution (elliptic flow $v_2$) must change at mid-rapidity from out-of-plane to in-plane emission~\cite{Ollitrault:1992bk} - a prediction confirmed at the BNL AGS and the CERN SPS. Prior to the start of the RHIC experimental programme, arguments based on fluid dynamics provided also some qualitative understanding of the dependence of elliptic flow on particle species, and on the energy and rapidity dependence of the collective sidewards displacement ($v_1$) of particle production at projectile 
rapidity~\cite{Appelshauser:1997dg}. Also, radial flow and femtoscopic data were 
interpreted at the time as resulting from a common, collectively expanding flow field, 
as formulated for instance in early 'blast wave models' 
(see Ref.~\cite{Wiedemann:1999qn} for a review). 

 The start of RHIC marked what I view as the first fluid dynamical
 revolution of heavy ion physics: The previously qualitative fluid dynamical picture of heavy ion collisions started to become quantitative.  This is largely a success of theory. 
 More precisely, it was clearly an important experimental achievement to 
 measure the $p_T$-differential elliptic flow $v_2(p_T)$ at RHIC with sufficient
 precision to identify a mild increase in comparison to flow measurements at lower
 center of mass energies. But that 
 the value of $v_2(p_T)$ comes close to saturating the maximal value of a perfect liquid was understood thanks to a theory effort into developing the tools for detailed fluid dynamic simulations~\cite{Teaney:2001av,Kolb:2001qz}.  On the basis of these numerical advances, 
 it was realized~\cite{Teaney:2003kp} that the approximate agreement 
 of $v_2$ with simulations based on a perfect liquid provides a tight constraint on the dimensionless ratio of shear viscosity over entropy density, $\eta/s$. 
 This quantity is of particular interest not only because it turns out to be smaller
 for the matter produced in heavy ion collisions than for any other known 
 fluid~\cite{Kovtun:2004de}. 
 More important is the theoretical insight that $\eta/s$ takes parametrically different values
 in different limiting formulations of collective motion, thus providing a highly sensitive
 test of the microscopic dynamics underlying fluid dynamic behavior. Namely, in weakly 
 coupled, HTL-resummed QCD perturbation theory, the normalized
 shear viscosity is parametrically large for perturbatively small coupling
 $\alpha_s$, $\eta/s \propto 1/\alpha_s^2 \log (1/\alpha_s)$~\cite{Arnold:2000dr}. 
 In contrast, Kovtun, Son and Starinets~\cite{Kovtun:2004de} (KSS) showed 
 that for a large class of non-abelian plasmas with gravity duals, 
 $\eta/s \to 1/4\pi \sim 0.08$ in the strong coupling limit ($g^2\, N_c \gg 1$). 
 Both, qualitative arguments~\cite{Kovtun:2004de} and exploratory 
 lattice calculations~\cite{Meyer:2007ic} suggest that $\eta/s$ in thermal QCD above 
 the transition temperature is close to this KSS limit.  Thus, by constraining $\eta/s \ll 1$
 from data~\cite{Song:2012ua}, we have learnt that the matter produced in heavy ion 
 collisions is a strongly coupled plasma with close to perfect fluid dynamical properties. 
 
 The KSS limiting value on shear viscosity marks also another aspect of the 
 first fluid dynamical revolution of heavy ion physics: string-theory based techniques 
 were recognized as a source of  guidance for heavy ion phenomenology. It is worth 
 recalling that QCD does not fall into the class of field theories with known gravity 
 dual. Rather, the gauge/gravity duality offers a framework for establishing rigorous 
 results for a large class of non-abelian plasmas that share many commonalities 
 with thermal QCD. To relate such results to heavy ion phenomenology requires 
 then an informed judgement of whether the result is a consequence of features 
 that can be expected to be more generic than the theoretical set-up in which they
 were obtained.  In the light of this caveat, it is truly remarkable to what extent the AdS/CFT
 correspondence has offered a framework for understanding central open questions
 in the phenomenology of heavy ion collisions~\cite{CasalderreySolana:2011us}. 
 Summarizing these successes lies beyond the scope of this conference talk, but I would
 like to highlight at least one case, the conceptual problem of early thermalization. Phenomenologically valid fluid dynamic simulations require early initialization 
 times $\tau_0 < 1$ fm/c, and thus early local equilibration, but to reach or maintain local 
 equilibrium, local collision rates must be much higher than the expansion rate 
 $\sim 1/\tau_0$ of the medium. At least for simple parametric estimates of the 
 collision rate in weakly coupled plasmas, $\sim \alpha_s^2\, T(\tau_0)$, 
 this condition cannot be realized for temperatures $T(\tau_0)$ expected at early times
 in a heavy ion collision. Although work continues on alternative, potentially more efficient 
 equilibration mechanisms in weakly coupled plasma
 (such as bottom-up thermalization~\cite{Baier:2000sb} or isotropization via plasma 
 instabilities), equilibration times $\tau_0 < 1$ fm/c remain difficult to reconcile 
 with a weakly coupled plasma.
 In contrast, it is by now convincingly demonstrated by use of methods in
 gauge/gravity duality that in the strong coupling limit of a large class of non-abelian
 quantum field theories at finite energy density,  generic far-from-equilibrium 
 distributions relax to close to thermal distributions on time scales 
 $\tau_0 \sim 1/T$~\cite{Heller:2011ju,Chesler:2009cy}. Thus, the paradigm
 of a strongly coupled plasma offers a natural underpinning of the two main assumptions
 on which fluid dynamical simulations are based: the assumption that a non-abelian 
 plasma of fundamental quantum fields can equilibrate fast and the assumption that
 it can expand almost isentropically, i.e. with minimal dissipative effects. 
 
 Moreover, while quasi-particle excitations are carried in weakly coupled
 plasmas over length scales $\tau_{\rm quasi} \sim 1/\alpha_s^2\, T$ that are large
 compared to the inter-particle distance $\sim 1/T$,  the strongly coupled 
 non-abelian plasmas formulated via gauge/gravity duality represent a dramatically different
 picture. These plasmas are unique in that they 
 do not carry quasi-particle excitations; the corresponding quasi-particle peaks in 
 spectral functions are missing. Since any quasi-particle excitation offers a channel
 to carry energy out of a locally equilibrated fluid element (and that means: to dissipate it), 
 a statement about the absence of quasi-particle excitations can be regarded as  
 explaining the microscopic dynamical origin of minimal dissipative effects 
 such as the ones measured by $\eta/s$. It is arguably one of the most
 fundamental open questions in heavy ion physics to establish up to which scale 
 this picture of a quasi-particle free microscopic structure of the plasma holds for thermal 
 QCD.  At this point, I recall only that the non-abelian plasmas formulated via 
 gauge/gravity duality correspond to conformal and thus scale-invariant quantum 
 field theories while the scale-dependence of QCD and asymptotic freedom implies 
 that at sufficiently high resolution scale the QCD plasma is seen as an ensemble of 
 quarks and gluons. Therefore, in contrast to the plasmas with known gravity duals, 
 thermal QCD may not be completely free of 
 quasi-particle excitations~\footnote{ Indeed, in quenched lattice QCD 
 calculations~\cite{Ding:2010ga}, quasi-particle peaks in some spectral functions are 
 seen to melt significantly but they do not disappear completely in contrast to 
 calculations based on the AdS/CFT correspondence.}. In short, the above arguments
 suggest that quasi-particle excitations represent an additional source of dissipation that 
 should lead to deviations from the minimal KSS bound and that may be present in QCD.
 Therefore, increasing the sensitivity of future determinations of $\eta/s$ to a level 
at which deviations of $\eta/s$ from the KSS limiting value become either visible or tightly
constrained could shed light on the microscopic nature of thermal QCD.

 Since QM2011 in Annecy, we are in what may be rightly called the second fluid dynamical
 revolution of heavy ion physics. It is based on the two basic insights that a fluid of minimal
 viscosity is a medium that has maximal transparency to primordial event-by-event
 fluctuations, and that the experimentally observed fluctuations are numerically large
 and thus accessible to detailed experimentation. So far, this second revolution is mainly experiment-driven. It starts with the finding that there are large harmonic coefficients 
 $v_n$ ($n$ odd) in the azimuthal Fourier decomposition of particle spectra. These 
 non-vanishing odd harmonics signal that the azimuthal symmetries of event-averaged nuclear 
 overlaps are broken on the level of single events. And as first stated by Alver and 
 Roland~\cite{Alver:2010gr},  the event-wise sampling of the collision geometry 
 induces event-by-event fluctuations with spatial anisotropies $\epsilon_n$ that 
 provide naturally the required symmetry breaking terms. As a consequence, the long-standing
 distinction in heavy ion physics between dynamical fluctuation measures and flow measures 
 seems to become obsolete at low $p_T$. Rather, a large class of measured 
 event-wise fluctuations at low $p_T$ is now understood as resulting from the fluid 
 dynamical propagation of initial fluctuations with event-wise primordial 
 fluctuations \cite{Holopainen:2010gz,Qiu:2011hf,Schenke:2011bn}
 and can thus be used for characterizing 
 fluid dynamical properties of thermal QCD. This picture received strong support from the 
 approximately linear relation between the $v_n$'s measured by all active experiments at 
 RHIC and at the LHC, and the spatial eccentricities $\epsilon_n$ (in particular $n = 2, 3$) 
 that were calculated for different models of initial conditions with event-by-event 
 fluctuation~\cite{Teaney:2010vd,Alver:2010dn,Bhalerao:2011yg}. But the 
 class of experimental measures of event-by-event fluctuations that can further test 
 the fluid dynamic picture of ultra-relativistic heavy ion collisions is significantly larger. 
 It includes knowledge of the probability distributions of the $v_n$'s in different centrality 
 classes~\cite{Gale:2012in}, correlations between the reaction planes $\Psi_n$ associated 
 to different flow harmonics~\cite{Jia:2012ju,Qiu:2012uy}, fluctuation measures with 
 particle-identification, etc.  
 
 There are simple parametric reasons for why the study of fluctuations will allow for precision
 measurements of properties of matter in the coming years. For instance, in a linearized
 treatment of fluid dynamic perturbations, excitations $\delta v(\tau,k)$ with 
 wave number $k$ decay exponentially on length scales set by the sound attenuation 
 length $\Gamma_s = \eta/(s\, T)$, namely $\delta v(\tau,k) =  \delta v(\tau_0,k) (\tau_0/\tau)^n 
 exp\left[-\Gamma_s\, k^2\, (\tau-\tau_0) \right]$, where the power $n$ depends on 
 the nature of the excitation, see e.g. \cite{Florchinger:2011qf}. On the one hand, 
 this illustrates that a 
 medium of minimal viscosity and hence minimal $\Gamma_s$ is indeed maximally 
 transparent to primordial fluctuations. On the other hand, it shows that an improved 
 control over the fluctuation scale $k$ is likely to result in an improved determination 
 of $\eta/s$.  
 And improving constraints on the phenomenologically dominant transport coefficient
 $\eta/s$ by a factor $\sim 5$, in addition to the motivations given above, has the 
 potential of making other dissipative properties, such as bulk viscosity and relaxation 
 times, experimentally accessible~\cite{Song:2012ua}. On the experimental side, 
 this argues in favor of
 exploiting all possibilities of controlling primordial fluctuations, including modifying 
 the shapes and sizes of the nuclear overlaps by varying beam species (such as in the recent
 exploratory Cu+Au or U+U runs at RHIC) or analyzing ultra-central 
 collisions~\cite{Luzum:2012wu} or 'engineering' event shapes~\cite{Schukraft:2012ah}. 
 On the theory side, there is of course a wide range of
 novel probes of fluid dynamic behavior that must be contrasted with fluid dynamical  
 simulations. But there are also important qualitative questions to be addressed. For
 instance, the strong $k$-dependence of the decay length of fluid dynamic excitations
 suggests that the characterization of
 primordial fluctuations may benefit from making their scale dependence more explicit
 than in the currently used parametrization in terms of spatial eccentricities $\epsilon_n$. 
 This may also help to provide support for the exciting possibility that the transverse scale of
 primordial fluctuations should be indicative of their quantum nature~\cite{Schenke:2012wb}, 
 rather than being
 characteristic of the nucleon diameter entering Glauber-type simulations (The latter 
 possibility would be difficult to understand dynamically. Despite the heuristic 
 use of the Glauber model for characterizing collision geometry, nucleons are unlikely 
 to be the relevant degrees of freedom for particle production at the TeV scale). 
 To mention but one other issue: there
 is no a priori reason for limiting a formulation of primordial fluctuations 
 to the transverse energy or entropy density. In principle, all fluid dynamic fields, 
 including transverse and longitudinal flow, can exhibit fluctuations at the initialization
 time $\tau_0$ of fluid dynamic simulations~\cite{Florchinger:2011qf}. 
 Given the rich phenomenology of fluctuation 
 measures, one may wonder to what extent such qualitatively different sources of primordial  fluctuations can be disentangled, and how this could contribute to a
 more precise understanding of initial conditions and their fluid dynamic propagation.
 
 The above argument that an improved fluctuation analysis gives access to detailed properties of matter is not unique to heavy ion collisions. It is for instance validated by the 
 recent successes of the Cosmic Microwave Background Cosmology that allowed
 to determine the material composition of our Universe with a few percent uncertainties
 from a fluctuation analysis in terms of multipole moments.  
 The analogies between these CMB fluctuation analyses and the analyses of heavy ion 
 collisions are indeed striking, and have been pointed out 
 repeatedly~\cite{Mishra,Mocsy,Naselsky:2012nw}.
 Here, I do not want to run through all aspects of this analogy. It is curious to note, however, that in the Big Bang studied in Cosmology, as well as in the Little Bangs 
 studied in heavy ion physics, fluctuations are propagated dynamically before 
 decoupling from the system, and that 
 it is the fluid dynamic nature of the propagation that gives access to material properties via the detailed analysis of fluctuations.  Moreover, in both cases, initial conditions 
 and their fluctuations are subject to significant uncertainties, but this did not prevent
 cosmology from entering a precision era. While one should be careful about 
 concluding too much from vague analogies between vastly different systems, 
 I expect that the analogy highlighted here will serve a dual purpose
 in the coming years: On the one hand, it will be interesting to assess to what extent analysis tools used in the CMB analysis  can be helpful in the fluctuation 
 analyses of nucleus-nucleus collisions. And on the other hand, the few \%  
 precision of key cosmological parameters achieved via CMB fluctuation analyses, 
 combined with the realization that a much lower precision is needed to address fundamental questions in heavy ion physics, may serve as a constant reminder that we are even in the soft physics sector far from having
 exploited the opportunities of experimenting with thermal QCD at RHIC and at the LHC.

\section{The opaqueness of non-abelian plasmas to hard probes}
As discussed above, a perfect liquid is a medium of maximal transparency to 
hydrodynamical excitations but it is a medium of maximal opaqueness to quasi-particle excitations. Compared to a statement about minimal shear viscosity, this latter 
statement is arguably a much more fundamental and generic assertion about the
properties of matter. However, at present it must be regarded as a hypothesis, 
conjectured solely on the basis of the observed perfect liquidity and on the basis of the 
properties of those theoretically explored strongly coupled non-abelian plasmas 
that are known to give rise to perfect liquidity. Can hard probes shed light on this hypothesis? How is the rather generic opaqueness of the medium to hard probes 
related to the properties of the perfect liquid? Will hard probes allow us to resolve
the microscropic partonic substructure of the medium that must emerge at sufficiently
high resolution scale in a plasma that knows about asymptotic freedom?
These are some of the broad range
of questions that we would like to address by utilizing hard probes as a tool for characterizing properties of matter. The numerically large medium modifications 
observed in almost all hard hadronic processes in heavy ion collisions give 
certainly strong support to the idea that hard probes are sensitive to the properties
of the produced medium. However, a detailed understanding of how this sensitivity
translates into statements about specific properties of matter is not yet achieved.
In the following, I shall discuss shortly how the properties of dense matter manifest
themselves in the propagation of hard probes within the different (simplified) theoretical
frameworks that have been developed for this question.

Let us discuss first what can be said about jet quenching in a perfect liquid,
if one takes the qualitative guidance from gauge/gravity duality at face value. How is the opaqueness of the medium to hard probes in this set-up related to the properties of the perfect liquid? In this framework, results are known for the 
propagation of several classes of hard probes~\cite{CasalderreySolana:2011us}.
In particular, light-flavored quarks or gluons injected into the plasma are known to
thermalize~\cite{Chesler:2008uy,Gubser:2008as}. More precisely, it has been shown
for sufficiently high quark energies $E \gg \sqrt{g^2 N_c} T$ that they 
have a maximal penetration depth~\cite{Chesler:2008uy}   $\Delta x_{\rm max}(E) \simeq
(1/2T) (E/T\sqrt{\lambda})^{1/3}$ beyond which they become indistinguishable from 
thermal fluctuations in the medium.
The situation is very different for heavy-flavored quarks,  where results are known for
sufficiently low quark energy, $\sqrt{E/M} \ll M/(\sqrt{\lambda} T)$~\cite{CasalderreySolana:2006rq,Herzog:2006gh}. Remarkably,
light-flavored quarks loose in a strongly coupled plasma most of their 
energy in the last stage of their propagation~\cite{Chesler:2008uy}, whereas the 
energy loss of heavy flavored quarks is proportional to their energy and occurs thus predominantly in the early stage. Moreover, because of flavor conservation, 
heavy-flavored quarks do not become indistinguishable from the thermal background. 
Rather, they thermalize only kinetically, subject to the random forces provided by the
perfect liquid. This process is governed by Langevin dynamics and the input parameters 
for this Langevin equation have been calculated by use of the gauge/gravity 
duality. The picture is therefore that heavy flavored quarks degrade their energy efficiently 
and that they will be dragged at late times with the flow field. 
Moreover, it is a direct consequence of a medium free of quasi-particle excitations that
the energy lost from hard probes is carried by hydrodynamic excitations, such as the
sound modes that interfere in a Mach cone, and the {\it stirring of the wake}~\cite{Gubser:2007ga}. 
There are continued efforts to relate data from RHIC and the LHC qualitatively or even
quantitatively to these and other results obtained within the framework of gauge/gravity duality~\cite{Horowitz:2012cf}. One may add the caveat here that even for bold 
attempts at doing phenomenology, exploratory calculations for hard probes 
in gauge/gravity duality need to be interfaced with further assumptions 
(such as assumptions about parton branching at late times outside the medium, hadronization, etc. ) that are difficult to control. This certainly complicates direct
attempts at assessing from data whether the mechanisms of parton energy loss 
identified in these calculations are realized in experimental reality. 

The calculations about the medium-dependence of hard probes summarized 
above are established firmly  in the strong coupling limit for 
non-abelian plasmas with gravity duals and that means: they are established
for scale-invariant theories. 
The coupling in these theories does not run, and irrespective of the momentum 
transferred between a probe and the perfect liquid, this probe will not resolve individual
partonic degrees of freedom in the plasma, simply because there is no scale at which
the perfect liquid can be viewed as being composed of them. In contrast, at 
sufficiently small scales, the QCD plasma should deviate from this picture of a perfect 
liquid because of asymptotic freedom. 
This qualitative idea of QCD-specific differences from a perfect liquid was
for instance nicely illustrated at this conference by a calculation of the probability distribution $P(k_\perp)$ of transverse momentum broadening of a
highly energetic parton in weakly coupled quark gluon plasma~\cite{D'Eramo:2012jh}. 
From gauge-gravity calculations, we know that this distribution is Gaussian with a 
width governed by the transport coefficient $\hat{q}$. The 
authors~\cite{D'Eramo:2012jh} showed in a HTL-resummed analysis to
leading order in the limit of large in-medium path length that $P(k_\perp)$ 
acquires a powerlaw tail $\propto 1/k_\perp^4$ at large transverse momenta, as 
expected for the rescattering of a partonic projectile on well-separated partonic
components in the medium. While such a dependence of $P(k_\perp)$ 
was realized previously in some models of medium-averages used in 
heavy ion collisions, this is the first model-independent field theoretic 
statement of this type.

This recent finding is but one illustration of the central question that is now coming 
into reach of jet quenching studies, namely the question of how jet quenching can inform us
about the microscopic constituents and macroscopic properties of the medium produced
in heavy ion collisions. For the example highlighted here, and in qualitative 
analogy to $\eta/s$ that distinguishes
via parametrically different values the cases of a perturbatively coupled and a strongly
coupled liquid, $P(k_\perp)$ at large $k_\perp$ distinguishes via parametrically
different values whether or not a hard probe views the medium as a collection 
of resolvable scattering centers. Given that $P(k_\perp)$ characterizes only one 
of several aspects of parton propagation in the medium, can $P(k_\perp)$ be 
determined experimentally, or are there experimentally better accessible quantities
to characterize via hard probes the microscopic structure of the plasma? At present, 
we do not have firm answers to questions of such microscopic nature. However,
as emphasized above, we know as a consequence of asymptotic freedom that
above an a priori unknown resolution scale, the jet interacts {\it perturbatively}
with resolved medium constituents even if the medium properties are 
largely governed by non-perturbative physics. Delineating the
kinematic regime in which a perturbative formulation of jet quenching applies, and 
understanding what the jet resolves in this kinematic range, has thus the potential to
teach us a lot about medium-properties. This is the basic motivation for developing
a perturbative formulation of jet quenching although the medium we are studying is
strongly coupled on soft scales. 
%
And to advance the interpretation of jet quenching 
data beyond the conclusions that were summarized at this 
conference~\cite{CasalderreySolana:2012bp,Milhano:2012cm}, one clearly needs
an improved control of how the results of perturbative parton energy loss calculations 
can be combined reliably in a medium-modified final state parton shower. 
In this context, I would like to highlight the formulation of jet quenching advanced by 
K. Zapp~\cite{Zapp:2011ek,Zapp:2012ak}, since it parallels closely the phenomenological
practice in high energy physics of modelling parton interactions by infrared continued pQCD
matrix elements supplemented with parton showers, and since it implements our current
knowledge of the non-Abelian LPM effect and collisional energy loss without any of the kinematical limitations in which analytical results on 'radiative' and 'collisional' energy loss
have been obtained so far. These features make the dominant uncertainties in this framework
quantifiable. Rather than entering at this stage a more detailed discussion of different jet quenching models, that lies clearly beyond the scope of this introductory talk, I would like
to end with an optimistic expectation: Formulating 
medium-modified final state parton showers with quantitatively controlled
uncertainties will parallel in the coming years the historical development of fluid dynamics,
where firm information about medium properties such as $\eta/s$ became also 
available only after having formulated these properties in a dynamically complete and quantitatively controlled framework.

I hope to have illustrated with these remarks on a small subset of important topics 
that both in theory and in experiment, and both
in the hard and in the soft physics sector, we are still 
far from having exploited the opportunities of accessing the microscopic structure
of thermal QCD produced at RHIC and at the LHC.

In preparing my talk at QM12 and in completing this presentation, I have profited
from discussions with many friends and colleagues. Here, I would like to single
out interactions with Stefan Fl\"orchinger, Ulrich Heinz, Krishna Rajagopal and 
Huichao Song.


\end{document}